\documentclass[reprint,superscriptaddress,showpacs,amsmath,amssymb,aps,prl,floatfix,lengthcheck]{revtex4-1}
\usepackage{graphicx}
\usepackage{dcolumn}
\usepackage{bm}
\usepackage{hyperref}

\begin{document}

\title{Infrared Vortex-State Electrodynamics in Type-II Superconducting Thin Films}
\author{Xiaoxiang Xi}
\affiliation{Department of Physics, University of Florida, Gainesville, Florida 32611, USA}
\affiliation{Photon Sciences, Brookhaven National Laboratory, Upton, New York 11973, USA}
\author{J.-H. Park}
\affiliation{National High Magnetic Field Laboratory, Florida State University, Tallahassee, Florida 32310, USA}
\author{D. Graf}
\affiliation{National High Magnetic Field Laboratory, Florida State University, Tallahassee, Florida 32310, USA}
\author{G. L. Carr}
\affiliation{Photon Sciences, Brookhaven National Laboratory, Upton, New York 11973, USA}
\author{D. B. Tanner}
\affiliation{Department of Physics, University of Florida, Gainesville, Florida 32611, USA}

\date{\today}

\begin{abstract}
The vortex-state electrodynamics of \textit{s}-wave superconductors has been studied by infrared spectroscopy. Far-infrared transmission and reflection spectra of superconducting Nb$_{0.5}$Ti$_{0.5}$N and NbN thin films were measured in a magnetic field perpendicular to the film surface, and the optical conductivity was extracted. The data show clear reduction of superconducting signature. We consider the vortex state as a two-component effective medium of normal cores embedded in a BCS superconductor. The spectral features are well explained by the Maxwell-Garnett theory. Our analysis supports the presence of magnetic-field-induced pair-breaking effects in the superconducting component outside of the vortex cores.
\end{abstract}
\pacs{74.78.-w, 74.25.Ha, 74.25.N-, 74.25.nd}
\maketitle

The vortex or Abrikosov state exists in type-II superconductors subjected to magnetic fields between $B_{c1}$ and $B_{c2}$. In this state the field penetrates the superconductor in the form of quantized tubes of flux, or vortices. The superconducting gap is zero inside the vortex cores and finite outside so that each vortex may be considered to have a core of normal metal, surrounded by superconductor~\cite{Abrikosov1957, Bardeen1965}. Because vortex quantization renders the material an inhomogeneous system, it necessarily affects the electrodynamics of the superconductor. The microwave response of the vortex state has been extensively studied theoretically \cite{Coffey1991,Brandt1991,Dulcic1993} and experimentally \cite{Belk1996, Powell1996, Sarti2005, Janjusevic2006, Zaitsev2007, Song2009}. However the picture is still incomplete in the infrared region spanning the superconducting gap \cite{Ikebe2009, Sindler2010}. 

In this Letter we address the infrared electrodynamics of the vortex state. We obtain the complex optical conductivity of type-II superconductors and compare our results to calculations of a superconductor-normal metal mixture using the two key models for the effective conductivity of an inhomogeneous system: that of Garnett
\cite{Garnett1905} (the so-called ``Maxwell-Garnett theory'' or MGT) and that of Bruggeman \cite{Bruggeman1935} (sometimes called the ``effective-medium approximation'' or EMA). We also compare our results to a theory of viscous motion of vortices driven by currents in the superconductor \cite{Coffey1991}. We find that only the MGT gives a good description of experiment, and then only when pair-breaking by the magnetic field \cite{AG1960,Skalski1964,Xi2010} is considered. That it does so is reasonable considering the topology of the vortex state: normal regions surrounded entirely by a connected superfluid. As pointed out some years ago \cite{Lamb1980}, this is the topology of the MGT: the inclusions are embedded in a host medium and are correlated to stay apart. In contrast, the EMA allows percolation of the minority constituent at some critical concentration, something that does not happen in the vortex state until the upper critical field, when the entire material is in the normal state.

\begin{figure}[b!]
\includegraphics[width=0.49\textwidth]{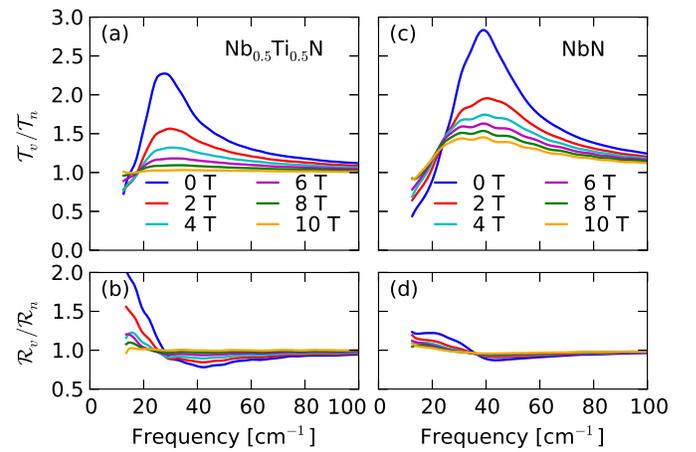}                
\caption{(color online) The vortex-state to normal-state transmittance ((a) and (c)) and reflectance ((b) and (d)) ratios for Nb$_{0.5}$Ti$_{0.5}$N and NbN at selected perpendicular magnetic fields. The reflection data include a stray-light correction discussed in the text.} 
\label{Fig1}
\end{figure}

\begin{figure*}[t!]
\includegraphics[width=0.95\textwidth]{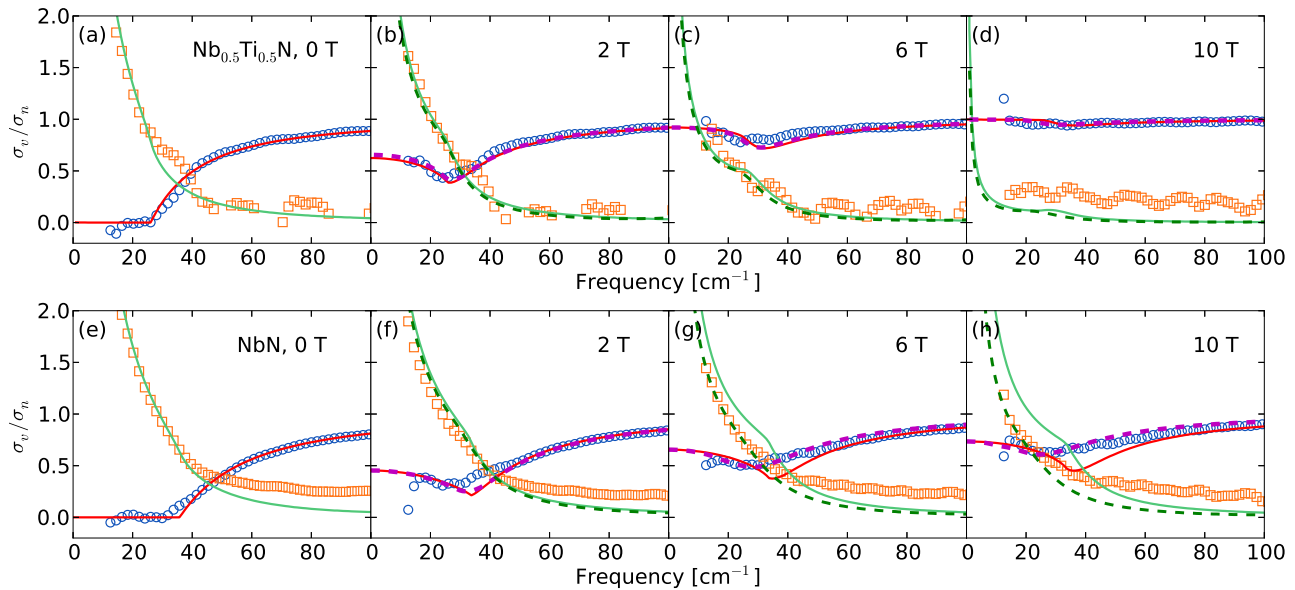}                
\caption{(color online) The real (circles) and imaginary (squares) part of the vortex-state optical conductivity $\sigma_v$ for Nb$_{0.5}$Ti$_{0.5}$N and NbN, normalized to the normal-state conductivity $\sigma_n$. The lines are MGT fits, neglecting (solid lines) and including (dashed lines) the pair-breaking effects.} 
\label{Fig2}
\end{figure*}

We studied type-II superconducting thin films of BCS superconductors Nb$_{0.5}$Ti$_{0.5}$N and NbN, which are widely used in superconducting magnets \cite{Kampwirth1985}, RF cavities \cite{Weingarten1996}, and photodetectors \cite{Guziewicz2011}. The 10~nm Nb$_{0.5}$Ti$_{0.5}$N film was grown on a quartz substrate in Ar and N$_2$ gas with a NbTi target, and the 70~nm NbN film grown on a MgO substrate in N$_2$ atmosphere using Nb, both by reactive magnetron sputtering \cite{Bosland1994,SuppMat}. The substrates have negligible absorption in the spectral range of interest (10--100~cm$^{-1}$) for $T<$~20~K. We performed the experiment at Beamline U4IR of the National Synchrotron Light Source, Brookhaven National Laboratory. The beamline is equipped with a Bruker IFS 66v FT-IR spectrometer, modified to use synchrotron radiation and a superconducting magnet for low-temperature magneto-spectroscopy. A composite silicon bolometer operating at $T\sim$1.5~K detects far-infrared radiation with high sensitivity. Both samples were cooled to 2~K ($\ll T_c$) in zero field, and their transmission and reflection measured in magnetic fields from 0--10~T, with the field direction normal to the sample surface. To avoid systematic experimental errors, the samples were brought to the normal state by heating to 20~K in zero field without changing the position, and their transmission and reflection measured. These transmission and reflection values served as a reference for calculating the real and imaginary parts of the optical conductivity relative to the normal state conductivity. An unavoidable reflection from a quartz vacuum window caused stray light to be included in reflection measurements \cite{SuppMat}. This was corrected by measuring the stray light and subtracting. Spectra were collected at 4~cm$^{-1}$ resolution, large enough to average out the interference fringes due to multiple internal reflections of the light in the substrate. The data for both samples at selected fields are shown in Fig.~\ref{Fig1}. As the magnetic field increases, the vortex-state transmission $\mathcal{T}_v$ and reflection $\mathcal{R}_v$ continuously approach the normal-state values $\mathcal{T}_n$ and $\mathcal{R}_n$. Nb$_{0.5}$Ti$_{0.5}$N reverts to the normal state more quickly than NbN due to its lower $B_{c2}$. The samples were also cooled through $T_c$ at each field shown in Fig.~\ref{Fig1}, and the same measurements repeated. The results show no noticeable difference from the zero-field-cooled case.

Measurements of both transmission and reflection enable the direct extraction of the thin film optical conductivity $\sigma = \sigma_1+i\sigma_2$. We measured the absolute normal-state transmittance $\mathcal{T}_n$, using an open aperture as reference. The films have large electronic scattering rates $1/\tau$ that $\sigma_n=\sigma_0/(1+i\omega\tau)\approx\sigma_0$ in the far-infrared, yielding a frequency-independent $\mathcal{T}_n$. From the measured $\mathcal{T}_n$ we extract $\sigma_0$, which in turn determines $\mathcal{R}_n$. Using the method described in Ref.~\cite{Xi2010}, we obtained $\sigma_1$ and $\sigma_2$ at various magnetic fields, shown respectively as circles and squares in Fig.~\ref{Fig2}. The extraction of $\sigma_2$ is prone to errors at high frequency, because it involves a small difference of two relatively large numbers. $\sigma_2$ is further complicated by the residual fringes in the raw data above 40~cm$^{-1}$. The optical conductivity data show two characteristic features: (i) With increasing magnetic field, the real part approaches the normal-state value (but does not exceed it whether below or above he gap frequency) while the imaginary part diminishes. (ii) A weak minimum in the real part appears around the gap frequency at finite fields.

We first compare these features with calculations based on two different effective medium theories for the average dielectric response of a mixture of two materials; the MGT and EMA \cite{Garnett1905,Bruggeman1935,Carr1985,SuppMat,Stroud1975}. The nanometer dimension of a vortex core is much smaller than the far-infrared wavelength. This justifies treating the vortex-state superconductor as an effective medium composed of a superconductor-normal metal mixture. The MGT and EMA make different assumptions, thus yielding different results. The MGT treats one constituent as the host and all others as embedded media, so is suitable for mixtures with isolated inclusions~\cite{Lamb1980}. The effective response functions vary smoothly with the volume fraction $f$ of the grains, and no percolation transition occurs for $f<1$. In contrast, the EMA treats all constituents equivalently, appropriate for mixtures with connected grains. It predicts percolation at a certain mixture fraction $f_c$. A vortex-state superconductor maintains its superconductivity even at a high volume fraction of vortices; percolation does not occur. Moreover, vortices correlate to stay apart because of the repulsive force between them. Therefore we expect the MGT to be a better description of our experiment, though one previous infrared study argued otherwise \cite{Sindler2010}. 

\begin{figure}[t]
\centering
\includegraphics[width=0.49\textwidth]{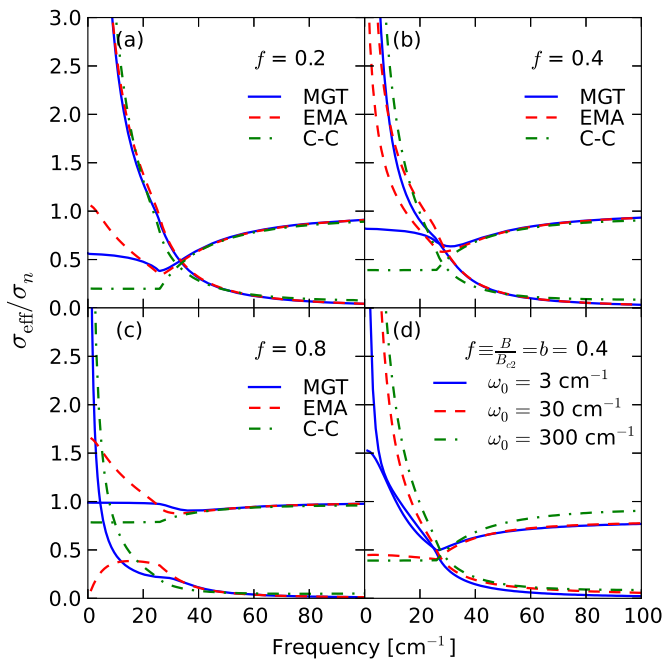}                
\caption{(color online) (a)--(c) Effective optical conductivity calculated from MGT (solid lines), EMA (dashed lines), and the Coffey-Clem model in the vortex-pinning regime at $\omega_0=$ 300~cm$^{-1}$ (dashed-dotted lines). $f$ is the normal-volume fraction. (d) Effective optical conductivity calculated from the Coffey-Clem model at different depinning frequencies $\omega_0$ for $f=0.4$.} 
\label{Fig3}
\end{figure}

Model calculation results in Fig.~\ref{Fig3} support the above reasoning. When calculating the effective optical conductivity, we treat the vortex cores as cylinders of dirty-limit Drude metal and the superconducting component using Mattis-Bardeen theory \cite{Mattis1958}. For the MGT the superconducting component is taken as the host medium. Fig.~\ref{Fig3}(a)--(c) compares the real and imaginary optical conductivity of the effective medium at various volume fractions $f$ of the normal metal. The MGT results are consistent with the experimental results shown in Fig.~\ref{Fig2}, including the minimum in the real conductivity near the gap frequency and the monotonic growth toward $\sigma_n$ for increasing field. The EMA results are strikingly different: the real conductivity is greater than $\sigma_n$ towards zero frequency, and changes non-monotonically with field. Moreover, the $1/\omega$ behavior of the imaginary conductivity is completely suppressed in the EMA for $f>0.5$, in contradiction to experiment (Fig.~\ref{Fig2}).

The effective medium theories discussed above assume a static vortex lattice. To consider the contribution of vortex motion, we compare our data to the model proposed by Coffey and Clem \cite{Coffey1991}, which calculates the surface impedance of type-II superconductors under the influence of vortex dynamics. They generalized the two-fluid model to couple self-consistently the supercurrent density with the vortex displacements. The central result of the model is a frequency-, field-, and temperature-dependent complex penetration depth $\tilde{\lambda}(\omega,H,T)$ \cite{Coffey1991,Coffey1993}. Noting the relation between the complex penetration depth and the complex optical conductivity $\sigma_{\mathrm{CC}}=i/\mu_0\omega\tilde{\lambda}^2$, and using the field-dependence of various quantities in Ref \cite{Coffey1993}, we obtain the complex optical conductivity in the low temperature limit ($T \ll T_c$),
\begin{equation}
\frac{\sigma_{\mathrm{CC}}}{\sigma_n} \approx \frac{(1-b)\sigma_s/\sigma_n +b}{\beta b(1-b)\sigma_s/\sigma_n+1}.
\label{sigma_CC}
\end{equation}
$b=B/B_{c2}$ is the reduced field. $\beta = 1/(1-i\omega_0/\omega)$, where $\omega_0$ is the characteristic frequency that distinguishes the flux-pinning and flux-flow regimes, called the depinning frequency. $\sigma_s$ is the optical conductivity of the superconducting component. Eq.~\eqref{sigma_CC} reduces to correct limits as $b\rightarrow 0$ and $b\rightarrow 1$. Since our superconducting samples were in the low temperature limit, we did not include the effect of thermal creep in deriving Eq.~\eqref{sigma_CC} and the expression for $\beta$.

Fig.~\ref{Fig3}(d) compares the effective Coffey-Clem optical conductivity calculated from Eq.~\eqref{sigma_CC} at different $\omega_0$. We assume $b\approx f$ to allow comparison with the calculation results of MGT and EMA, also shown in the figure. (The validity of this approximation will be discussed below.) We note that using $\omega_0=300$~cm$^{-1}$ corresponds to the flux-pinning regime and $\omega_0=3$~cm$^{-1}$ the flux-flow regime. In the flux-pinning regime, the real conductivity agrees with the data as well as with MGT and EMA above the gap, but almost flattens below the gap. Such behavior misses the minimum around the gap, found at different values of $f$ in Fig.~\ref{Fig3}(a)--(c) and in the data in Fig.~\ref{Fig2} (b,c,d,f,g,h). Flux flow brings a minimum to the real conductivity Re($\sigma_{\mathrm{CC}}$), but it also substantially steepens the low-frequency part of Re($\sigma_{\mathrm{CC}}$) and suppresses the high-frequency part. These features render the Coffey-Clem theory inconsistent with experiment.

Having established that the MGT possesses the salient features of the experimental data, we now focus on a quantitative analysis. Previous studies of the vortex-state electrodynamics in superconductors assumed the superconducting fraction to be unaffected by the magnetic field \cite{Dulcic1993,Ikebe2009,Sindler2010}. Adopting this assumption, we use the Mattis-Bardeen theory \cite{Mattis1958} to describe the superconducting fraction and the Drude model for the vortices. The volume fraction of the normal vortex cores $f$ is varied to fit the real optical conductivity data at different fields. Once an optimal fit is found, we use $f$ to calculate the imaginary part. The fits and calculations are shown as solid lines in Fig.~\ref{Fig2}. The zero-field case is trivial as we set $f=0$. The minimum in the real conductivity around the gap frequency is captured by the fits, but appears shallower in the data than in the fits. This indicates less missing spectral weight in the data, meaning weaker superconductivity than assumed for the fits. The low-frequency imaginary conductivity also corroborates this view. The calculated imaginary conductivity, especially that of NbN, is above the data as the field increases. Therefore, the analysis points to additional weakening of superconductivity beyond the presence of the vortex cores.

One source of such additional weakening could be magnetic-field-induced pair-breaking effects in the superconducting fraction. This effect needs to be considered because the magnetic field outside of the vortex cores does not diminish significantly until several penetration depths away from the core center \cite{Brandt2002}. The vortex spacing for a square lattice \cite{Sonier2004} is $a_{\square}=\sqrt{\Phi_0/B} \approx 49$~nm at 1~T and less at higher fields, where $\Phi_0$ is the magnetic flux quantum and $B$ is the applied field. These distances are smaller than the penetration depth of our samples (a few hundred nm). Therefore the field penetrates all of the superconducting component for our measured field range. This field breaks the time-reversal symmetry of the electron pairing, resulting in pair-breaking effects \cite{AG1960,Skalski1964}. Such effects reduce the energy gap and smear out the gap-edge singularity in the quasiparticle density of states, consequently modifying the optical conductivity. These effects have been demonstrated in our studies of Nb$_{0.5}$Ti$_{0.5}$N and NbN thin films in parallel magnetic fields \cite{Xi2010,Xi2012}.

We re-evaluate the MGT analysis by including the pair-breaking effects. A complete model requires computing the field distribution in the superconducting fraction, and then the pair-breaking optical conductivity as a function of distance from the vortex center. For simplicity, we use the pair-breaking optical conductivity from our parallel-field study. Such an approximation is valid when the average field in the superconducting fraction is close to the applied field. In fact, for the field-normal-to-film geometry, taking into account a demagnetization factor $\sim$1, the average internal field almost equals the applied field \cite{Doria2008}. Taking the value of $f$ from the previous fits, we calculate the effective optical conductivity using MGT. The results are shown in FIG.~\ref{Fig2}(b)--(d) and (f)--(h) as dashed lines. Inclusion of the pair-breaking effects significantly improves the quality of the fits, especially for NbN at high fields where such effects become significant. The calculated optical conductivity, both the real and imaginary parts, matches better to the data. The improvement is hardly noticeable for Nb$_{0.5}$Ti$_{0.5}$N mainly because its $B_{c2}$ at 2~K is about 11~T; thus the greater density of vortex cores dominates the average response, leaving the pair-breaking effects difficult to see. 

\begin{figure}[t]
\centering
\includegraphics[width=0.35\textwidth]{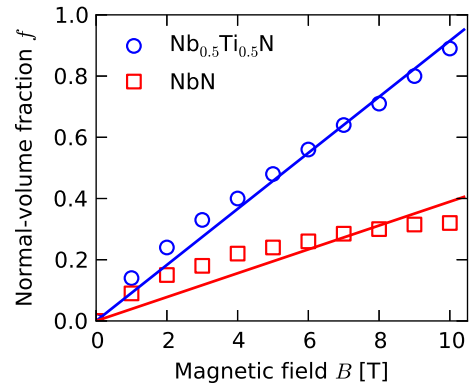}                
\caption{(color online) Field dependence of the normal-volume fraction $f$ for Nb$_{0.5}$Ti$_{0.5}$N (circles) and NbN (squares). The solid lines are fits to $f=B/B_{c2}$.} 
\label{Fig4}
\end{figure}

The analysis presented above yields the normal-volume fraction $f$. Its value at different fields for both samples is plotted in Fig.~\ref{Fig4}. We previously assumed $f =B/B_{c2}$ when discussing the Coffey-Clem model. This assumption can be justified by noting that $f=\pi r_v^2/a_{\square}^2=\pi r_v^2 B/\Phi_0\propto B$, where we have assumed a square vortex lattice with lattice spacing $a_{\square}$ and vortex core radius $r_v$. A linear fit yields $B_{c2}=10.9$~T for Nb$_{0.5}$Ti$_{0.5}$N and $B_{c2}=25.6$~T for NbN. For comparison we determined $B_{c2}$ of both samples by four-probe resistivity measurements at Station SCM2 at the National High Magnetic Field Laboratory \cite{SuppMat}. We found $B_{c2}\approx 11$~T for Nb$_{0.5}$Ti$_{0.5}$N and $B_{c2}>20$~T for NbN. While the two methods give consistent result, we point out that the field dependence of $f$ shown in Fig.~\ref{Fig4} is sub-linear for both samples. Such behavior is accounted for by the shrinking of the vortex cores with increasing field. Field-induced reduction of the vortex core radius by more than a factor of two has been observed in conventional \textit{s}-wave superconductors \cite{Sonier2004}, explained in terms of either the ``vortex lattice squeezing effect'' or the inter-vortex transfer of bounded quasiparticles in the vortex cores. Although the latter effect was argued to be significantly weakened by scattering \cite{Kogan2005}, it has been observed in dirty Pb thin films \cite{Ning2010}. When the vortex core radius $r_v(B)$ decreases with field, the normal-volume fraction $f=\pi r_v^2(B) B/\Phi_0$ could become sub-linear in field.  

In conclusion, we measured the far-infrared transmission and reflection of Nb$_{0.5}$Ti$_{0.5}$N and NbN thin films in perpendicular magnetic fields, and observed superconductivity weakening. The extracted optical conductivity of the vortex state is consistent with the Maxwell-Garnett theory for normal metal inclusions in a superconducting host. Detailed analysis suggests that in the vortex state the magnetic field weakens superconductivity via at least two ways. One is the increasing normal metal component as the vortex density increases with field. The other is the weakening of the BCS state in the remaining superconducting component due to pair-breaking effects. We did not need to include the flux-flow-induced dissipation in our analysis. Effects of flux flow are expected to be important in cleaner materials with less pinning.
\vspace{6mm}

We are grateful to S. W. Tozer for the access to the equipment used in the four-probe resistivity measurements, and to G. Nintzel and T. P. Murphy for technical assistance. Work at University of Florida, Brookhaven National Laboratory and National High Magnetic Field Laboratory was supported by U.S. Department of Energy through DE-FG02-02ER45984, DE-ACO2-98CH10886 and DE-FG52-10NA29659.

\clearpage

\setcounter{figure}{0}
\setcounter{table}{0}
\setcounter{page}{1}
\pagenumbering{arabic}
\noindent
\textbf{Supplemental Material}

\section{I. S\lowercase{ample} P\lowercase{arameters}}
The sample parameters relevant to our study are listed in Table~\ref{table1}. The film thickness $d$ was estimated from the film growth conditions. The zero-temperature gap $\Delta_0$ was extracted from the zero-field optical conductivity, shown in Fig.~2 of the main text. The critical temperature $T_c$ was determined from temperature-dependent resistivity measurements. The normal-state sheet resistance $R_{\square}$ was determined from the normal-state transmittance, 
\begin{equation}
\mathcal{T}_n = \frac{4n}{(n+1+Z_0/R_{\square})^2},\tag{S1}
\end{equation}
where $n$ is the substrate refractive index and $Z_0$ is the vacuum impedance. We found the penetration depth $\lambda$, coherence length $\xi$, and the substrate refractive index $n$ from literature \cite{NbTiN_prm,NbN_prm,refr_idx}. 
\begin{table}[b]
\renewcommand{\thetable}{S\arabic{table}}
\centering
\caption{Sample parameters} \label{table1}
\begin{tabular}{cccccccc}
\hline
Sample &   $d$  & $\Delta_0$ & $T_c$ &  $R_{\square}$  & $\lambda$ & $\xi$ & $n$\\
       &(nm) & (cm$^{-1}$)   &  (K)  &  ($\Omega$)  &  (nm) & (nm)  & \\\hline
Nb$_{0.5}$Ti$_{0.5}$N  & 10 & 12.8 & 10.2 & 117  & 200--400 & 3.8--5.0   & 2.12\\ 
NbN    & 70 & 17.9 & 12.8 & 48   & 180--500 & 4.0--7.0  & 2.90\\ \hline
\end{tabular}
\end{table}

\begin{figure*}[b]
\renewcommand{\thefigure}{S\arabic{figure}}
\includegraphics[width=0.95\textwidth]{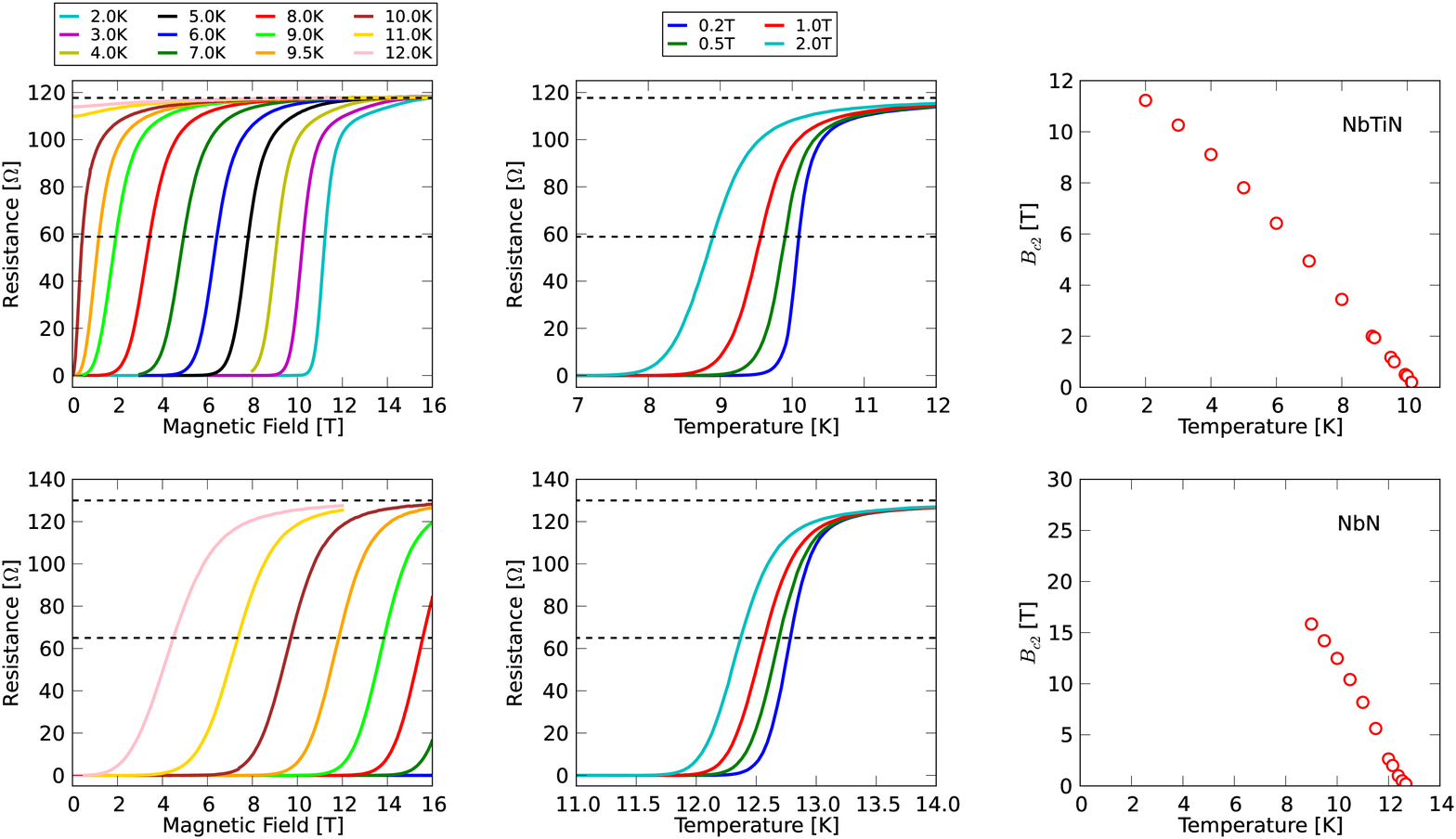} 
\caption{Four-probe resistance data in perpendicular fields for Nb$_{0.5}$Ti$_{0.5}$N (first row) and NbN (second row). Left column: resistance vs perpendicular field at different temperatures. Middle column: resistance vs temperature at different perpendicular fields. Right column: circles are $B_{c2}$ extracted from data. In the resistance plots dashed lines indicate the resistance at 20~K and half of that value.}  
\label{FigS1}
\end{figure*}

To determine the upper critical field $B_{c2}$ of both samples, we performed four-probe resistivity measurements at Station SCM2 at the National High Magnetic Field Laboratory. A 2~mm$\times$2~mm piece was cut from each sample for such measurements because of the requirements of the experimental set-up. Electrical contacts were made by gluing gold wires to the thin films with silver paint. The two samples were mounted simultaneously on a rotating probe, which allows accurate alignment of the sample surface with respect to the magnetic field orientation. Resistivity was measured with the field normal to the film surface. Well below the critical temperature, the resistance was measured by scanning the magnetic field between 0~T and 16~T at selected temperatures. Close to the critical temperature, the resistance was measured by sweeping temperature between 5~K and 20~K at selected low field values. The data are shown in Fig.~\ref{FigS1}. Because the transition is broad, we define $B_{c2}$ as the field at which the resistance drops to 1/2 of its 20 K value. We determined $B_{c2}\approx 11$~T at 2~K for Nb$_{0.5}$Ti$_{0.5}$N. $B_{c2}$ at 2~K for NbN is beyond the measurement range of the magnet. Based on the temperature-dependent trend shown in Fig.~\ref{FigS1}, we estimate that it should be greater than 20~T at 2~K.

\section{II. I\lowercase{nfrared} M\lowercase{easurement} P\lowercase{rocedures}}
\begin{figure*}
\renewcommand{\thefigure}{S\arabic{figure}}
\includegraphics[width=0.8\textwidth]{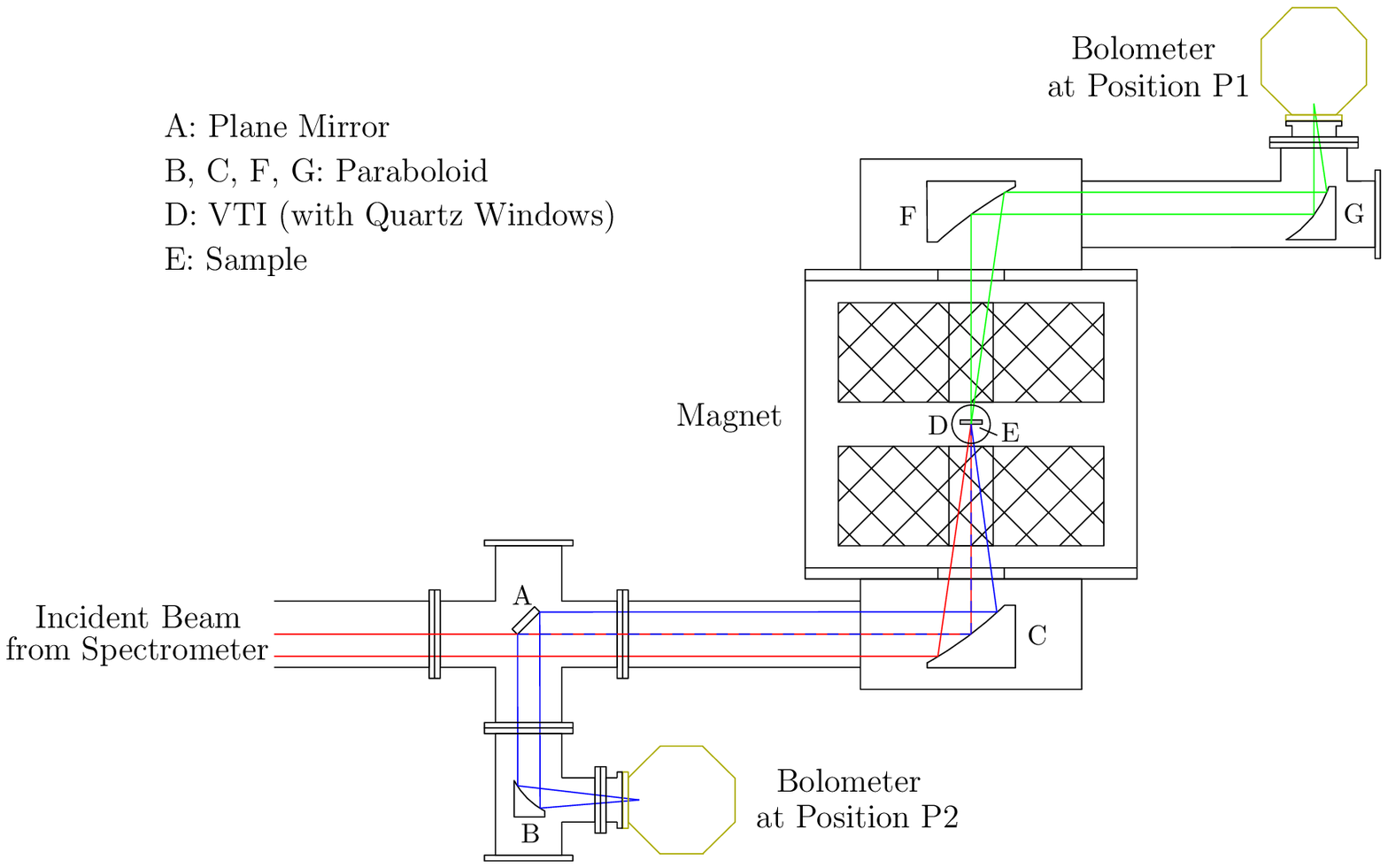} 
\caption{Optical layout for transmission and reflection measurements in a perpendicular magnetic field. The red rays are the incident beam, the green rays the transmitted beam, and the blue rays the reflected beam. A plane mirror is installed in the four-way cross-shaped flange to direct beams in the reflection measurement, and is taken out to allow the full beam to pass in the transmission measurement. The bolometer is placed at Position P1 for transmission and moved to Position P2 for reflection measurement.}
\label{FigS2}
\end{figure*}

For transmission and reflection measurements, the samples were loaded into the superconducting magnet at the bottom of a probe in the Faraday configuration (external magnetic fields perpendicular to the film surface). We started with transmission measurement. In the experiment layout shown in Fig.~\ref{FigS2}, the plane mirror (A) for collecting the reflected light was taken out from the four-way cross-shaped flange, so that the incident beam could completely pass through. The transmitted signal was detected by the bolometer placed at Position P1. To apply the field perpendicular to the film, we first roughly loaded the samples in the approximate angle range, and then carefully rotated the probe to maximize the transmission signal. The position was carefully marked for later reference. The error of the alignment is expected to be within 5$^{\circ}$. 

The reflection measurement was done immediately after the transmission measurement. The sample was kept at the same position, with the orientation of the sample probe unchanged so that the magnetic field was perpendicular to the film surface. Plane mirror A was installed to direct the reflected signal from the sample to the bolometer, which had been moved from Position P1 to Position P2 downstream of the plane mirror. There are two sets of quartz windows, one set on the exterior of the magnet, the other set on the Variable Temperature Insert (VTI) in which the sample was mounted. The stray light was mainly from the entrance quartz window on the VTI. This stray light was measured by rotating the sample probe 45$^{\circ}$ from the marked position mentioned in the previous paragraph, and subtracted from all the single beam reflection spectra of the samples. To minimize the stray light, the plane mirror (A) and the paraboloid (B) were manipulated to minimize the ratio of the stray signal and the signal from the sample at the marked position (true signal plus stray signal).

\section{III. E\lowercase{ffective} M\lowercase{edium} T\lowercase{heories}}
Here we derive the effective optical conductivity of a vortex-state superconductor from the Maxwell-Garnett theory and the Bruggeman effective medium approximation. The basic formulation of effective medium theories can be found in literature, e.g. \cite{Carr1985S}. 

Consider an inhomogeneous medium made of two components, grain $a$ with volume fraction $f$ embedded in the surrounding $b$ with volume fraction $1-f$. The electromagnetic waves traversing the medium can be treated as a spatially-averaged field with its electric component expressed as
\begin{equation}
\langle \mathbf{E}\rangle = f\mathbf{E}_a + (1-f)\mathbf{E}_b.\label{eq_Eave}\tag{S2}
\end{equation}
The response function $\mathbf{D}$ is similarly spatially averaged,
\begin{equation}
\langle \mathbf{D} \rangle = f\epsilon_a \mathbf{E}_a +(1-f)\epsilon_b \mathbf{E}_b.\label{eq_Dave}\tag{S3}
\end{equation}
The effective dielectric function is then 
\begin{equation}
\epsilon_{\text{eff}} = \frac{\langle\mathbf{D}\rangle}{\langle\mathbf{E}\rangle}.\label{eq_epseff}\tag{S4}
\end{equation}

\subsection{1. Maxwell-Garnett theory}
The Maxwell-Garnet theory considers the electric field $\mathbf{E}_a$ to be the local field acting on the grain $a$. It consists of the external field and the field caused by the polarized charges on the surface of an artificial cavity in which the grain resides. In considering this local field, the theory assumes that the separation between grains are large enough so that individual grains scatter light independently. Furthermore, the field in the surrounding medium $b$ is assumed to be unaffected by the presence of the grains \cite{Troy1999S,Granqvist1977S}. Based on these assumptions, the Maxwell-Garnett theory gives the effective dielectric function for oriented ellipsoid grains as 
\begin{equation}
\epsilon_{\mathrm{MGT}} = \epsilon_b+\epsilon_b\frac{f(\epsilon_a-\epsilon_b)}{g(1-f)(\epsilon_a-\epsilon_b)+\epsilon_b},\tag{S5}
\label{eq_MGT_epsilon}
\end{equation}
where $f$ is the volume fraction of grain $a$, and $g$ is the depolarization factor determined by the shape of the ellipsoid. To convert to the optical conductivity, note $\epsilon = 1+ 4\pi i\sigma/\omega$, and $\sigma_a=\sigma_n$ and $\sigma_b = \sigma_s$. Vortices are generally regarded as cylindrical tubes with a normal core of radius of the coherence length $\xi$, each carrying a quantum of magnetic flux $\Phi_0$. Therefore we set $g=1/2$.

\subsection{2. Bruggeman effective medium approximation}
\label{sec_EMA}
Bruggeman proposed a method to address the issue that the grains and host material in Maxwell-Garnett theory are treated asymmetrically. Because of the presence of grains with different properties from the surrounding medium, the electric field in the region around the grains is modified, and the electric flux deviates from that when such grains are absent. Bruggeman argued that the average flux deviation for the whole medium should vanish \cite{Troy1999S}. He then suggested that an adequate choice of a self-consistent local field can satisfy this condition. This leads to the consideration of an effective medium in which all inclusions are treated on an equal basis, and the average flux deviation is zero \cite{Granqvist1977S}. Equivalently, the effective dielectric function can be calculated using Eqs.~\eqref{eq_Eave},~\eqref{eq_Dave}, and~\eqref{eq_epseff}. The difference from the Maxwell-Garnett theory is that there is no host medium in Bruggeman effective medium approximation. Both $\mathbf{E}_a$ and $\mathbf{E}_b$ are calculated assuming the constituents $a$ and $b$ are grains immersed at their related volume fractions in an effective medium characterized by an effective dielectric function.

For an inhomogeneous medium made of two components, grain $a$ of volume fraction $f$ and grain $b$ of volume fraction $1-f$, Bruggeman effective medium approximation gives the effective dielectric function for oriented ellipsoid grains as the solution of the following equation,
\begin{equation}
f\frac{\epsilon_a-\epsilon_{\mathrm{EMA}}}{g\epsilon_a+(1-g)\epsilon_{\mathrm{EMA}}}+(1-f)\frac{\epsilon_b-\epsilon_{\mathrm{EMA}}}{g\epsilon_b+(1-g)\epsilon_{\mathrm{EMA}}}=0.\tag{S6}
\label{eq_EMA}
\end{equation}
Treating the vortices as cylindrical particles, we set $g=1/2$. The solution is 
\begin{align}
\epsilon_{\mathrm{EMA}} &= \frac{1}{2}\bigg[(2f-1)(\epsilon_a-\epsilon_b)+\nonumber\\
&\sqrt{(2f-1)^2(\epsilon_a-\epsilon_b)^2+4\epsilon_a\epsilon_b}\bigg].\tag{S7}
\end{align}
Only this one out of the two solutions of the quadratic equation~\eqref{eq_EMA} is chosen, because it reduces to the correct values in the limit $f\rightarrow 0$ and $f\rightarrow 1$. The optical conductivity is obtained using the relation $\epsilon = 1+ 4\pi i\sigma/\omega$, and $\sigma_a=\sigma_n$ and $\sigma_b = \sigma_s$.

\end{document}